# Real-time gait event detection using motion capture to control an electrical stimulator: Proof-of-concept

Graffagnino G[1], Gasq D[2,3], Patte K[4], Sijobert B[4], Azevedo-Coste C[1]

[1]Inria, University of Montpellier, France
[2]Toulouse NeuroImaging Center, University of Toulouse, France
[3]Service des Explorations Fonctionnelles Physiologiques, Clinique Universitaire du Mouvement, Centre Hospitalo-Universitaire de Toulouse, France
[4]Institut St Pierre, Palavas-les-Flots, France

***Abstract:*** Cerebral palsy (CP) is the most prevalent motor disorder in childhood and often results in gait abnormalities that hinder mobility and diminish quality of life. Functional electrical stimulation (FES) has demonstrated potential in enhancing gait in individuals in this population, however, its practical implementation remains complex, as it requires monitoring various gait parameters and delivering personalized stimulation to different muscles in order to correct various gait impairments. Recent advancements in real-time motion capture (MOCAP) and wearable sensors now enable the development of closed-loop, multi-channel FES systems. This study will assess the feasibility and responsiveness of a real-time, event-triggered multi-channel stimulation protocol during treadmill walking. The stimulation is triggered by specific gait events (heel strike, knee flexion, ankle dorsiflexion) detected through the MOCAP system and administered via a multichannel electrical stimulator. Conducted on healthy adults, this preliminary study focuses on assessing technical feasibility. We report different technical outcomes including the latency between gait event detection and the appearance of stimulation artifacts in EMG signals. The results confirm the viability of the system, laying the groundwork for future clinical application in the rehabilitation of children with CP.

## Introduction

Cerebral palsy (CP) is the most prevalent motor disorder in childhood, affecting approximately 1 in every 500 live births worldwide. It involves a spectrum of motor impairments - such as increased muscle tone, muscle weakness, and poor coordination - that significantly limits functional mobility and reduce quality of life [1]. Gait abnormalities are among the most common functional limitations observed in individuals with CP, often resulting in reduced walking efficiency, impaired balance, and limited independence [2]. Recent advances in wearable sensors and real-time motion analysis have opened new avenues for the development of closed-loop electrical stimulation techniques, which adapt stimulation in response to specific biomechanical events. In this context, functional electrical stimulation (FES) has emerged as a promising method to support motor function by enhancing muscle activation in a task-specific manner [3]. Over the past two decades, single-channel gait-specific FES has been shown to improve ankle motion during the swing phase but remains insufficient for addressing more complex gait impairments, such as flexed-knee or stiff-knee gait, commonly observed in CP [4]. Emerging evidence suggests that multi-channel gait-specific FES may provide more comprehensive gait correction, though further research is needed to validate its effectiveness [5]. On the other hand, real-time detection of gait events and real-time feedback emerge as a potential new way of improving rehabilitation of children with cerebral palsy, using motion capture [6]. As motion capture can visualize various kinematic parameters, we aimed to combine real-time detection of multiple specific gait events using motion capture to control multi-channel gait-specific FES to address a new functional tool of gait correction of children with cerebral palsy. Our approach would be to determine, in a gait analysis laboratory, the most relevant events to detect in order to trigger stimulation of various muscles, to test different stimulation strategies, and then to transfer the selected one to a personalized wearable system [7]. To advance this line of research, it is essential to first evaluate the feasibility and reliability of such systems in a controlled environment.

The present study aims to evaluate from a technical perspective, a real-time, event-triggered electrical stimulation protocol during treadmill walking in healthy individuals, as a first step toward validating its feasibility and effectiveness for future applications in children with CP.

## Material and Methods

The objective of this study is to compute the latency between the detection of various gait events and the reception of a stimulation on the muscles of interest. The study design is schematized in figure 1.

**Participants:** 10 participants: from 18 to 60 years old, without any neurologic disorder, able to walk on a treadmill without any assistance.

**Electrical Stimulation:** Electrical stimulation is delivered by the 8 channel Motimove stimulator (3F Company, Belgrade, Serbia), operated in research mode via hexadecimal command signals. Stimulation frequency and phase width were set to 30Hz and 150us respectively.

**Data collection:** Kinematic and electromyographic (EMG) data were collected using the Gait Real-time Analysis Interactive Lab (GRAIL, Motek Medical, Houten, Netherlands) and processed through the D-Flow software, in which a custom interactive task was developed in accordance with the study design. Twenty-two reflective markers were placed on anatomical landmarks of both legs and tracked at 100 Hz using a 10-camera optoelectronic motion capture system (Vicon Motion Systems Ltd., Oxford, UK), employing the Human Body Model (HBM) for real-time gait analysis, as described by Van den Bogert et al. [8]. Stimulation artifacts were recorded using a multi-channel acquisition system (Cometa Systems, Bareggio, Italy) at a sampling rate of 1000 Hz.

**Study design:** Participants were instructed to walk normally on a treadmill across three separate sessions. In the first session, the focus was on detecting the heel strike event; each time a heel strike is identified, electrical stimulation is triggered. The second session involved real-time monitoring of knee flexion for each leg. When the knee angle exceeds a predefined threshold, stimulation is triggered. The third session followed a similar protocol, but based on ankle dorsiflexion to trigger stimulation.

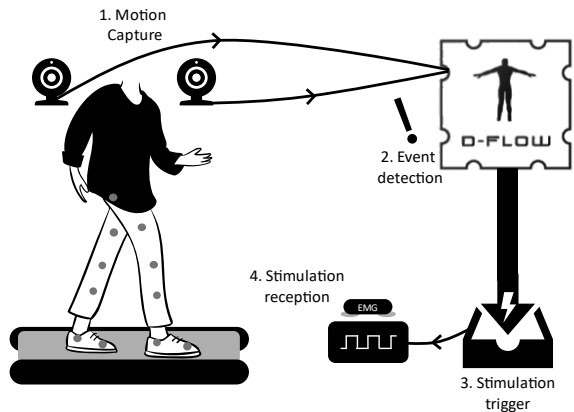

Figure 1: Schema of study design

**Data processing:** All data collected using the D-Flow software were imported and processed in MATLAB (v2020b, The MathWorks Inc., Natick, USA). The timestamps of specific gait events (heel strike, knee flexion, ankle dorsiflexion) were extracted and compared to the timestamps of the first appearance of stimulation artifacts and the stimulation duration in the EMG signals. This analysis allows for the determination of the delay between event detection and the actual delivery of stimulation to the body. The measured latency was then compared to the duration of each corresponding gait event to assess whether the delay is acceptable for real-time gait correction, or if an anticipatory stimulation offset is required to ensure effective intervention.

## Results

The results demonstrate the feasibility of triggering electrical stimulation with a latency of less than 100 ms using the motion capture system integrated within the GRAIL platform. Details will be presented during the conference.

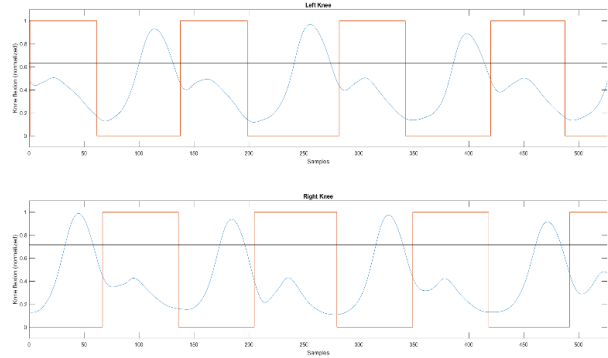

Figure 2: Example of the experiment on the knee flexion. In blue knee flexion normalized to maximal knee flexion. In orange, stimulation activation (on/off). In black, knee flexion threshold (stimulation is on when flexion is below this threshold).

## Discussion

The primary objective of this study was to assess the latency of stimulation delivery based on gait events, these results were combined with previously published electromechanical delay values [10] to estimate the total latency of our corrective application.
Healthy participants exhibit typical gait patterns, which may not pose the same challenges to the stimulation system as pathological gait. This limits the generalizability of latency and responsiveness outcomes in more variable or unpredictable gait conditions. The stimulation was applied below the motor threshold, and we will need to verify the adaptation to an FES-assisted gait. In parallel to this work, we have assessed the event detection approach on retrospective data acquired on the GRAIL platform with children with CP.

## Conclusions

This study proposed to evaluate a real-time, event-triggered electrical stimulation protocol during treadmill walking in healthy adults, as a preliminary step towards gait correction in children with cerebral palsy. The system leverages motion capture technology to detect biomechanical events and trigger multi-channel stimulation with minimal latency. While stimulation was delivered below the motor threshold, the recorded delay—combined with known electromechanical response times—offers insight into the feasibility of timely muscle activation for gait modulation. Although conducted in a controlled environment and limited to a healthy population, the results provide a foundational understanding of the system's responsiveness and technical reliability. This work lays the groundwork for developing adaptive, sensor-based neurorehabilitation strategies tailored to the complex motor impairments characteristic of cerebral palsy.

## Acknowledgement

This work was funded by an INRIA-INSERM grant. We gratefully acknowledge the Physiotherapy Department and the management of the St Pierre Institute for granting access to the GRAIL system and supporting the data acquisition for this protocol.

## Author's Address

GRAFFAGNINO Gabriel
CAMIN Team, Inria, University of Montpellier, France
gabriel.graffagnino@inria.fr